\def\figcond{1}     
\def\figs#1#2#3#4#5#6{%
\begin{figure}[tbhp]
\centerline{\hbox{\hsize=5in\hss\vbox to #2in{\vss%
\ifnum\figcond>0
\vskip #4in%
\centerline{\hskip #3in\epsfig{file=#5,width=#1in,angle=90}}
\else
\centerline{$#5$}
\fi
\vss}\hss}}
\caption{#6}
\end{figure}}
\ifnum\figcond>0
\documentstyle[preprint,floats,prc,aps,epsfig]{revtex}
\else
\documentstyle[preprint,floats,prc,aps]{revtex}
\fi
\begin{document}

\preprint{DOE/ER/40427-02-N95}
\draft

\title{QUARK MODEL DESCRIPTION OF THE N-N SYSTEM:\\
       MOMENTUM DISTRIBUTIONS, STRUCTURE FUNCTIONS AND THE EMC EFFECT}

\author{W. Koepf}

\address{School of Physics and Astronomy\\
Raymond and Beverly Sackler Faculty of Exact Sciences\\
Tel Aviv University, 69978 Tel Aviv, Israel}

\author{L. Wilets}

\address{Department of Physics, FM-15\\
University of Washington, Seattle, WA 98195, USA
\bigskip}

\date{February 26, 1995}

\maketitle

\begin{abstract}

We employ a relativistic quark bag picture, the chromo-dielectric
soliton model, to discuss the quarks' symmetry structure and momentum
distribution in the $N$-$N$ system. Six-quark clusters are constructed
in a constrained mean-field calculation. The corresponding
Hamiltonian contains not only an effective interaction between the
quarks and a scalar field, which is assumed to parameterize all
non-perturbative effects due to the non-linearity of QCD, but also
quark-quark interactions mediated through one-gluon-exchange.  We
also evaluate the quark light-cone distribution
functions, characterizing inclusive deep-inelastic lepton scattering,
for the nucleon as well as for the six-quark structures.
We find a competition between a softening of the quarks' momenta
through the increase of the confinement volume, and a hardening via
the admixture of higher symmetry configurations due to the
color-electrostatic one-gluon-exchange.  These findings suggest an
unexpected absence of many-nucleon, multi-quark effects, even though
six-quark structures should represent a non-negligible part of the
nuclear ground state.

\end{abstract}
\bigskip

\section{INTRODUCTION}

In a previous investigation \cite{koepf94}, the
nucleon-nucleon interaction in terms of quark degrees of freedom
was studied.
Employing a relativistic quark bag model, the
chromo-dielectric soliton model (CDM) \cite{fai88},
six-quark bags  were constructed, representing two nucleons at various
distances from complete separation to total overlap.
Through a constrained mean-field calculation, molecular-type
configurations \cite{stancu87} were generated as a
function of deformation, and the configuration mixing was evaluated
in a coupled channel analysis.
The effective Hamiltonian
contained not only a scalar background field, but also quark-quark
interactions mediated through one-gluon-exchange.
The result was
an adiabatic nucleon-nucleon potential which was repulsive at short
distances, and the origin of which was attributed to the
color-electrostatic one-gluon-exchange.
In this work, we  focus our
attention on the quark sub-structure of the $N$-$N$ system, employing
the six-particle wave functions obtained previously.

Well over ten years ago, it had been suggested that the nucleon's
radius swells in the nuclear medium, and that overlapping
three-quark structures thus have lower momenta than well separated
nucleons \cite{noble81}.  It had furthermore
been claimed \cite{pirner81}, that the clustering of nucleons into
six-quark structures can account for the EMC effect \cite{emc}, i.e.,
the change of the nucleon's properties in the nuclear environment
as observed in deep-inelastic scattering of highly
relativistic leptons off nuclei, if one assumes that
the quarks' momentum distribution within those clusters is
shifted to lower momenta \cite{brodsky74}.

In the following, we  address those issues by means of a detailed
analysis of the quark sub-structure of the $N$-$N$ state, within
the model description presented before. We investigate the
quark's momentum distribution in  six-quark clusters as a
function of the inter-nucleon separation, and show that there
exists an interesting competition between a softening of the quark's
momenta through the increasing confinement volume
and a hardening via the admixture of higher configurations,
which are usually neglected in investigations
in that realm, and that had proven to be important
in the study of the $N$-$N$ potential.

We furthermore evaluate the deep-inelastic structure functions
of the nucleon as well as of the six-quark configurations,
following the formalism presented by the Adelaide group
\cite{signal88} for the calculation of the nucleon's
structure functions from the MIT bag model \cite{mit}.  In this
description, which has since been extended to other
non-topological soliton models \cite{bate92}, the twist-two
light-front quark distribution functions are calculated from the bag
model wave functions at an, {\it a priori}, unknown renormalization scale,
including a Peierls-Yoccoz projection \cite{peierls57} to cure the broken
translational invariance and form momentum eigenstates.
In order to compare with
data, the calculated quark distribution functions are
then evolved from the relatively low scale, at which
the bag model is expected to be a reasonable approximation to the
non-perturbative QCD bound state, to the
experimental momentum scale using the non-singlet
Altarelli-Parisi evolution equations \cite{altarelli77}.

The outline of this work is as follows.  In Sec. II, we review
the model with which we describe the $N$-$N$ system in terms of quark
degrees of freedom.  In Sec. III we discuss the symmetry structure
and quark momentum distribution of the $N$-$N$ ground state
for various inter-nucleon separations.
In Sec. IV, we present the formalism used to
calculate the deep-inelastic structure functions for
the nucleon, and in Sec. V we extend this technique
to the six-quark case.  Finally, we summarize and conclude in Sec. VI.

\section{THE MODEL}

The Lagrangian of the chromo-dielectric, non-topological soliton
model \cite{fai88} is the fundamental QCD
Lagrangian supplemented by a scalar field, which
parameterizes the bulk of the non-perturbative effects
which arise due to the non-linearity of QCD, and which
simulates the gluonic condensate and other scalar structures that
inhabit the physical vacuum.  The model Lagrangian,
\begin{equation}
{\cal L}~=~\overline{\psi} i \gamma^\mu D_\mu \psi
{}~+~\textstyle{{1 \over 2}} \partial_\mu \sigma \partial^\mu \sigma
{}~-~ U(\sigma) ~-~\textstyle{{1 \over 4}} \kappa(\sigma)
F_{\mu\nu}^c F^{\mu\nu c} \ ,
\end{equation}
is covariant and satisfies chiral symmetry. Here,
\begin{equation}
U(\sigma)~=~{a \over 2!}\sigma^2~+~{b \over 3!}\sigma^3~ +~{c \over
4!}\sigma^4~+~B\
\end{equation}
is the self-interaction energy of the scalar field, $F_{\mu\nu}^c$
is the color-$SU(3)$ gauge field tensor, and
\begin{equation}
\kappa(\sigma)~=~1~+~\theta(\sigma)\left({\sigma\over\sigma_v}
\right)^{\!2}
\left[\, 2 \, {\sigma\over\sigma_v} \, - \, 3 \, \right]
\end{equation}
is the color-dielectric function, where $\sigma_v$ is the scalar
field's vacuum expectation value.

Although there is no direct quark-sigma coupling, the quarks still
acquire a self-energy through their interactions with the
gluon field, and this leads to spatial confinement \cite{fai88}.
Color confinement,
on the other hand, arises through the enclosure of the quark cavity
by the physical vacuum where the dielectric function goes to zero.
The latter also ensures that there are no spurious color Van der
Waals forces \cite{vdw}, which trouble many non-relativistic
investigations in that realm.

In order to fit the parameters of the model, we construct
self-consistent solutions for the nucleon.  To simulate spatial
confinement, we add an effective
coupling between the quarks and the scalar field,
\begin{equation}
{\cal L}_{q\sigma}~=~-~g_{eff}(\sigma)~\overline{\psi}~\psi\ ,
\end{equation}
with
\begin{equation}
g_{eff}(\sigma)~=~g_0~\sigma_v \left(~{1 \over
\kappa(\sigma)}~-~1~\right) \ .
\end{equation}
We employ the coherent state approach and treat the scalar field
classically.  In addition, gluonic terms are dropped when
determining the quark wave functions or the scalar field, and the
N-$\Delta$ mass splitting is used to adjust the strong coupling
constant, $\alpha_s$.

The six-quark system is then investigated by means of a constrained
mean-field calculation for various ``inter-nucleon distances" from
total separation to complete overlap.  Through an external
potential \cite{schuh86}, quark wave functions are generated as a
function of a geometric deformation parameter, $\alpha$, which for
large positive values coincides with the true inter-nucleon
separation, for $\alpha = 0$ corresponds to a spherical bag and for
negative $\alpha$ to oblate deformations.  We limit ourselves to the
lowest states of either parity, $|\sigma\rangle$ and $|\pi\rangle$
in the molecular notation of Ref. \cite{stancu87}, with magnetic
quantum numbers of $m=\pm 1/2$.

Quark-gluon interactions are treated in the one-gluon-exchange
(OGE) approximation, as higher order effects are assumed to be
simulated by the scalar field. This yields Abelian gluon field
equations,
\begin{equation}
\partial^\mu \Bigl(~\kappa(\sigma)~\bigl[ \partial_\mu A^c_\nu -
\partial_\nu A^c_\mu \bigr]~\Bigr)~=~
{g_s \over 2} \, \overline \psi \gamma_\nu
\mbox{\boldmath $\lambda$}^c \psi \ .
\end{equation}
Note that the gluon field is explicitly affected by the scalar
background through the color-dielectric $\kappa(\sigma)$, and the
gluonic propagators are thus evaluated ``in medium"
\cite{bicke85}.  We choose the Coulomb gauge,
$\mbox{\boldmath $\nabla$}\cdot [\kappa(\sigma)~{\bf A}^c]=0$,
to deduce the mutual and self-interaction terms,
and finally calculate their contribution to the one-
and two-body parts of the
effective Hamiltonian.  The corresponding matrix elements are
evaluated by means of ``fractional parentage coefficients"
\cite{harvey81}, and they generate an explicit mixing between the
various six-quark configurations.
The part of the OGE interactions that arises from the time
component of the gluonic field is responsible for
color confinement, and the part of the OGE interactions that stems
from the spatial components
generates the color-magnetic hyperfine interaction which, in turn,
produces the N-$\Delta$ mass splitting.

In Ref. \cite{koepf94} results were presented for the adiabatic,
local $N$-$N$ potential, obtained in Born-Oppenheimer approximation
from the energy difference of a deformed six-quark bag and
two well separated, non-interacting nucleons.  The
isospin-spin
channels (TS)=(01) and (TS)=(10), which are compatible with $L=0$
partial waves, were studied for two different
parameter sets, denoted as $f=3$ and $f=\infty$, adjusted to the
standard properties of the nucleon. A purely repulsive
central interaction was found with a ``soft" core whose
maximum varied between 200 and 350 MeV for the two sets.
The long and medium range attraction could not be reproduced since it
should be attributed to explicit meson exchange and not quark
rearrangement, and the ``sea" quarks were not included.
Also, the non-gluonic contribution of the Hamiltonian was
attractive, while the OGE part was dominantly repulsive. In
particular, it was the color-electric one-gluon-exchange, mediated
by $A_0^c$, which lead to the repulsion at small $N$-$N$ separations,
and not the spin-spin color-magnetic hyperfine interaction, which in
the literature is quoted \cite{barnes93} as being responsible for the
short-range repulsive core.  For further details we refer the reader
to Ref. \cite{koepf94}.  We note that the dynamics of the
$N$-$N$ interaction have been shown to contribute to the repulsion
as well \cite{schuh86}.

\section{SYMMETRY STRUCTURE OF THE SIX-QUARK STATE}

The construction of antisymmetric six-quark
states is a central part of any study of the $N$-$N$ system in
terms of quarks.  Incorporating all possible degrees
of freedom -- color (C), orbital motion (O), spin (S) and isospin (T)
-- we employ a classification scheme based on $SU(4)$ spin-isospin
symmetry, as introduced by Harvey \cite{harvey81}.  For the orbital
share of the wave function, we use ``molecular orbitals"
\cite{stancu87} expressed through spatial single-particle states that
are wave functions of a static Hamiltonian, and
which in our case are the two lowest orbitals of either parity,
$|\sigma\rangle$ and $|\pi\rangle$.  If we furthermore restrict
ourselves to the isospin-spin channels (TS)=(01) and (TS)=(10), only
the following seven configurations, displayed in terms of the Young
tableaux characterizing their permutation symmetries,
\begin{eqnarray}
|1\rangle~&=&~|NN\rangle\ ,                          \nonumber \\
|2\rangle~&=&~|\Delta\Delta\rangle\ ,                \nonumber \\
|3\rangle~&=&~|CC\rangle\ ,                          \nonumber \\
|4\rangle~&=&~|42^+ [6]_O~ [33]_{TS} \rangle\ ,      \\
|5\rangle~&=&~|42^+ [42]_O [33]_{TS} \rangle\ ,      \nonumber \\
|6\rangle~&=&~|42^+ [42]_O [51]_{TS} \rangle\ ,      \nonumber \\
|7\rangle~&=&~|51^+ [6]_O~ [33]_{TS} \rangle\ ,      \nonumber
\end{eqnarray}
contribute significantly to the $N$-$N$ ground state.  The first
three contain solely configurations
which asymptotically are of the type $|R^3 L^3\rangle$, and
the other four are of the form $|R^4 L^2+R^2 L^4\rangle$ or
$|R^5 L+R L^5\rangle$, denoted as $42^+$ and $51^+$ respectively,
where, for general deformation, $|R\rangle$ and $|L\rangle$ are
replaced by $|r\rangle$ and $|\ell\rangle$ with
\begin{equation}
|r,\ell\rangle~\equiv~{|\sigma\rangle \pm |\pi\rangle \over \sqrt{2}}
\ .
\end{equation}
The pseudo-right/left states, $|r\rangle$ and $|\ell\rangle$,
approach $|R\rangle$ and $|L\rangle$ asymptotically, and the latter
correspond to a quark centered at the location of
either one of the two respective
nucleons.  All the states in Eq. (7) have color symmetry
$[222]_{C}$, and they mix through the one-gluon-exchange interaction.

\figs{11.75}{3.0}{0.3}{-2.2}{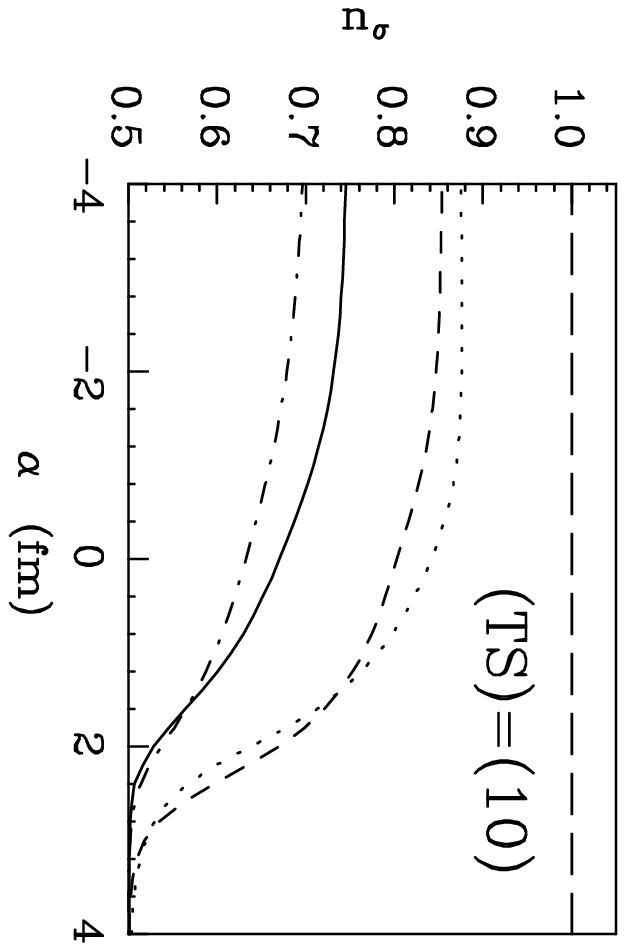}{The occupation probability of
the positive parity state in the $N$-$N$ clusters
as a function of deformation.  The short-dashed (dotted)
lines show results of calculations where the color-magnetic hyperfine
interaction was included, the dot-dashed (solid) lines correspond to
calculations where also the color-electrostatic force was included,
and the long-dashed line refers to a calculation where the OGE was
left out altogether.  Results are shown for (TS)=(10), and for the
parameter sets $f=\infty$ ($f=3$).}

Na\"\i vely one would expect that once the individual bags overlap
and the degeneracy between the single-particle states vanishes,
all quarks occupy the lowest spatial orbital,
and this hypothesis is an implicit ingredient
to all works where a lowering of the quarks' momenta in
six-quark clusters is discussed \cite{pirner81}.  To test that
assumption, we show in Fig. 1 the probability of the quarks
in the $N$-$N$ clusters to occupy the lowest single-particle state,
$|\sigma\rangle$,
\begin{equation}
n_\sigma~=~{\textstyle{1 \over 6}}~\langle \, 6q \, | \,
{\hat a}_\sigma^\dagger \, {\hat a}_\sigma \, | \, 6q \, \rangle \ ,
\end{equation}
where the six-quark ground state
was obtained while including different parts of the
one-gluon-exchange contribution in the diagonalization of the
effective Hamiltonian.

The quantity $n_\sigma$ is shown as a function of the parameter
$\alpha$, which characterizes the deformation of the external
potential used to generate the single quark wave functions,
and which for large positive $\alpha$ coincides with the true
inter-nucleon separation, $r_{NN}$.  Note that the spherical
configuration, $\alpha=0$, corresponds to a still finite
inter-nucleon separation, and that $r_{NN} \to 0$ is approached
for oblate deformations, i.e., for $\alpha<0$.

We observe that only if the gluonic interaction
is neglected completely can all quarks move
into the lowest single-particle orbital.  The inclusion
of the one-gluon-exchange, and the channel-coupling it generates,
leads to a non-negligible occupation of higher states, of the form
$|\sigma^4\pi^2\rangle$ and $|\sigma^2\pi^4\rangle$, in particular
if not only the color-magnetic hyperfine but also
the color-electrostatic interaction is considered.  The latter is
frequently neglected in investigations in that realm, and it had
proven to be essential for obtaining the short-range repulsion in
the investigation of the $N$-$N$ interaction in Ref. \cite{koepf94}.
Actually, neglect of the gluonic interactions in a {\it dynamical}
calculation would lead to $n_\sigma=1/2$.  This is because
$n_\sigma=1/2$ for well separated nucleons, i.e., the $|NN\rangle$
state, and only gluonic interactions (or Coriolis mixing) could
change the occupation of the individual orbitals in, e.g., a
time-dependent mean-field calculation.

In Fig. 2 we show the single quark's average momentum,
\begin{equation}
\langle k \rangle~\equiv~{\textstyle{1 \over 6}}~\sqrt{\langle \,
6q \, | \, {\hat k}^2 \, | \, 6q \, \rangle} \ ,
\end{equation}
as a function of the deformation of the six-quark clusters, and again
employing different shares of the one-gluon-exchange
contribution to the effective Hamiltonian.
If the one-gluon-exchange is neglected and all quarks are assumed to
occupy the lowest single-particle orbital, i.e.,
$|6q\rangle\sim|\sigma^6\rangle$, the increase in the size of the
confinement volume leads to a strong decrease of the quark's momenta
as soon as the nucleonic bags overlap considerably.  As mentioned
earlier, this observation is at the heart of the numerous
investigations where either the swelling of the nucleon in the nuclear
medium \cite{noble81} or the contribution of many-nucleon, multi-quark
effects to nuclear observables is discussed \cite{pirner81}.

\figs{11.75}{5.2}{0.3}{0.3}{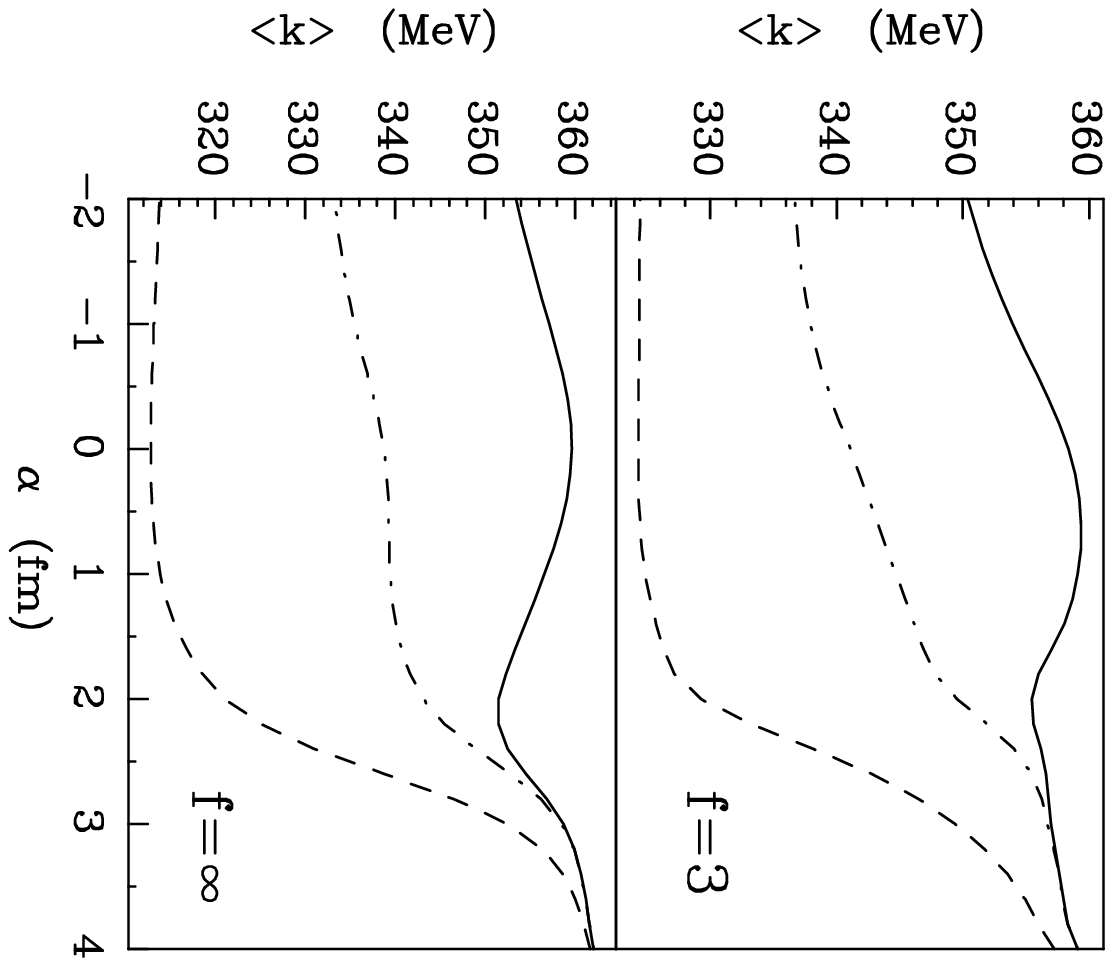}{The average quark momentum
in the six-quark clusters as a function of deformation.
Results are shown for (TS)=(10) and for the two parameter
sets $f=3$ and $f=\infty$. The dashed lines correspond to
calculations where the one-gluon-exchange was left out and all
quarks were assumed to occupy the lowest orbital,
the dot-dashed lines show the results of calculations where
the color-magnetic hyperfine interaction was included,
and the solid lines refer to calculations where also
the color-electrostatic interaction was accounted for.}

However, the inclusion of the
one-gluon-exchange interaction leads to an admixture
of higher states, as has been depicted in Fig. 1, and hence also
to a hardening of the quarks' average momentum.  In particular,
when not only the color-magnetic hyperfine interaction but
also the color-electrostatic share of the OGE is considered,
the aforementioned hardening almost counterbalances
the softening from the increase in the size of the confining
volume.  This cancellation points towards an unexpected absence of
multi-quark effects, even if six-quark structures represent a
non-negligible part of the nuclear ground state.

\section{DEEP-INELASTIC STRUCTURE FUNCTIONS}

As one the most prominent areas in which six-quark effects are
supposed to play a dominant role \cite{pirner81} is the EMC effect,
i.e., the change of the properties of the nucleon in the
nuclear environment as observed in deep-inelastic lepton scattering,
we evaluate in the following the unpolarized,
deep-inelastic structure functions of the nucleon as well as of
the six-quark clusters.

\figs{9.0}{2.0}{1.9}{2.3}{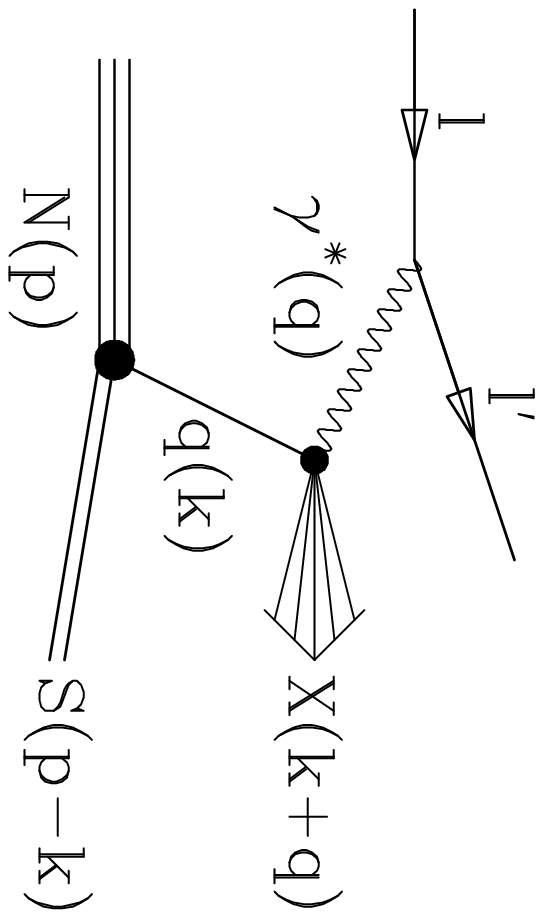}{Inclusive scattering of a
high-energy lepton from a nucleonic target. The virtual photon,
$\gamma^*$, is absorbed by a quark, q, in the target nucleon, N,
leaving behind a spectator, S, and producing an undetected final
state, X. The quantities in brackets are the particles' momenta.}

Consider the inclusive scattering of a high-energy lepton from a
nucleonic target, as depicted in Fig. 3.  We assume that a
spacelike virtual photon is absorbed by a quark with momentum $k$ in
the target nucleon, leaving behind a ``spectator" and producing an
arbitrary undetected final state. In the laboratory frame, the
relevant momenta are
\begin{mathletters}
\begin{eqnarray}
q~&=&~\left( \, \nu \, , \, {\bf 0_{\bot}} \, , \, -|{\bf q}| \,
\right) \ , \\
p~&=&~\left( \, M \, , \, {\bf 0} \, \right) \ , \\
k~&=&~\left( \, k_0 \, , \, {\bf k} \, \right) \ ,
\end{eqnarray}
\end{mathletters}
where $M$ is the nucleon mass.
It is customary to characterize the spacelike photon by
means of its virtuality, $Q^2 \equiv |{\bf q}|^2 - \nu^2$, and
through the Bjorken scaling variable, $x_B \equiv {Q^2 \over 2M\nu}$.
In the following, we are only interested in the kinematic limit
where $Q^2 \rightarrow \infty$ while $x_B$ remains finite.
Transforming into the Breit (or infinite-momentum) frame, where the
energy of the exchanged virtual photon vanishes, we find to leading
order in an expansion in $1/Q^2$,
\begin{mathletters}
\begin{eqnarray}
q'~&=&~\left( \, 0 \, , \, {\bf 0_{\bot}} \, , \, -Q \, \right) \ , \\
p'~&\approx&~\left( \, {Q^2+2M^2x_B^2 \over 2Qx_B} \, , \, {\bf 0_{\bot}}
\, , \, {Q \over 2x_B} \, \right) \ , \\
k'~&\approx&~\left( \, {Q^2(k_0+k_z)+2M^2x_B^2k_0 \over 2MQx_B} \, , \,
{\bf k_{\bot}} \, , \, {Q^2(k_0+k_z)+2M^2x_B^2k_z \over 2MQx_B} \,
\right) \ .
\end{eqnarray}
\end{mathletters}
The quark's light-cone
momentum fraction $x$ is defined
as the ratio of the $\hat z$-components of the momenta of the struck
quark and the target nucleon in the Breit frame, i.e.,
\begin{equation}
x~\equiv~{k'_z \over p'_z}~\approx~{k_0+k_z \over M}~\approx~x_B \ ,
\end{equation}
where the second and third equalities are valid to leading order in
$1/Q^2$ only.

According to Jaffe \cite{jaffe83}, the leading-twist contribution to
the quark distribution functions,
which characterize the probability to find a quark of flavor $f$
carrying a fraction $x$ of the target's light-cone momentum,
may be written as
\begin{equation}
q_f(x)~=~\sum_S~\delta \left(x-k_+/M\right)~| \langle \, S \, | \,
{\hat \psi}_{+,f} (0) \, | \, N \, \rangle |^2 \ ,
\end{equation}
where the sum runs over all
possible intermediate spectator states, and
where $k_+$ is the plus-component of the struck quark's momentum
in the laboratory frame, $k_+=k_0+k_z$.  Employing this
equation, the nucleon's structure functions can, in principle,
be calculated from any
phenomenological, low-energy model. The respective formalism was
developed for a evaluation of the nucleon's structure functions from
the MIT bag model \cite{signal88}, and it has
since been extended to other non-topological soliton models
\cite{bate92}, and to the nuclear structure
functions \cite{saito92,naar93}.
In the sum in Eq. (14), we restrict ourselves to a single two-quark
intermediate spectator state, which we furthermore assume to
be on its mass-shell with mass $M_S$.  This, in turn, fixes the
struck quark's energy to $k_0 = M-\sqrt{M_S^2 + |{\bf k}|^2}$.
Obviously, the truncation of the sum in Eq. (14) after only one term
is a very strong approximation, and we continue to discuss
its validity in the remainder of this section.

This then yields
\begin{equation}
q_f(x)~=~{1 \over (2\pi)^3}~\sum_m~\langle \, \mu \, | \, P_{f,m} \,
| \, \mu \rangle~ \int d^3{\bf k} ~{\phi_2(|{\bf k}|) \over \phi_3(0)}
{}~\left| \psi_{+,f,m} ({\bf k}) \right|^2 \ ,
\end{equation}
where $|\mu\rangle$ is the $SU(6)$ spin-flavor wave function of the
initial nucleon, $P_{f,m}$ is a projector onto flavor $f$ and
helicity $m$, and
\begin{equation}
\left| \psi_{+,f,m} ({\bf k}) \right|^2~=~\delta \left(x-k_+/M\right)
\int d^3{\bf r} \int d^3{\bf r'}~\psi^\dagger_{f,m}({\bf r})~
{1+\alpha_z \over 2}~
\psi_{f,m}({\bf r'})~e^{i {\bf k} ({\bf r} - {\bf r'})}
\end{equation}
is the Fourier transform of the twist-two contribution to the
connected matrix element of the current-current correlator.
The $\delta$-function, which fixes the
plus-component of the struck quark's momentum, can be rewritten as
\begin{equation}
\delta \left(x-k_+/M\right)~=~{M \over |{\bf k}|}~\delta\!\!\left(
\cos\theta_k -
{\sqrt{M_S^2 + |{\bf k}|^2} - M(1-x) \over |{\bf k}|} \right) \ ,
\end{equation}
and it yields a lower limit in the integration over the struck
quark's three-momentum, i.e., $|{\bf k}| \geq k_{min}$, where
\begin{equation}
k_{min}~=~\left| {M^2(1-x)^2 - M_S^2 \over 2M(1-x)} \right| \ .
\end{equation}

{}From this we observe that the quark distribution has
support not only for $0 \leq x \leq 1$, but also for negative $x$.
However, as shown by Jaffe \cite{jaffe83}, there are other
semi-connected diagrams,
which are not included into our formalism, and which exactly
cancel the contributions for negative $x$. We
hence proceed and simply neglect that spurious domain.

In Eq. (15) we also included a Peierls-Yoccoz
\cite{peierls57} projection onto momentum eigenstates,
expressed by means of the Hill-Wheeler overlap functions,
\begin{equation}
\phi_n(|{\bf k}|)~=~4\pi \int_0^\infty dR~R^2 j_0(|{\bf k}|R)~T^n(R)
\ ,
\end{equation}
with
\begin{equation}
T(R)~=~\int d^3{\bf r}~\psi^\dagger_{f,m}({\bf r})~
\psi_{f,m}({\bf r}+{\bf R}) \ ,
\end{equation}
and where $n=2$ for the intermediate di-quark state, and $n=3$ for
the initial nucleon.

As noticed first by Close and Thomas \cite{close88}, the mass of the
spectator state depends on whether the latter
is in a spin singlet or in a spin triplet, due to the
one-gluon-exchange. This is usually built into model descriptions
of that type by assigning different masses, $M_S^{S=0}$ and
$M_S^{S=1}$, to the intermediate di-quark singlet and triplet states,
and it has been demonstrated \cite{signal88} that this yields a
satisfactory description of the quark distribution functions'
flavor and spin dependence.  As in this work, we are only
interested in flavor and spin averaged distributions,
we neglect this effect in the following, and simply set
\begin{equation}
M_S~=~M - \epsilon \ ,
\end{equation}
where $\epsilon$ is the quark's single particle energy,
which has the values $\epsilon=340$ MeV ($f = 3$) and
$\epsilon=335$ MeV ($f = \infty$) for the two parameter sets.

Previously, we constructed self-consistent solutions for the
nucleon.  We used those calculations to
fit the parameters of our model, $a$, $b$ and $c$ in $U(\sigma)$ of
Eq. (2), $g_0$ in $g_{eff}$ of Eq. (5) and the
strong coupling constant $\alpha_s$, and we presented
solutions for $f \equiv b^2/ac = 3$ and $f = \infty$.
In the following, we evaluate Eq. (15) with the respective
$s_{1/2}$ quark wave functions,
\begin{equation}
\psi_m({\bf r})~=~\left(\begin{array}{c}
f(r) \\ i \mbox{\boldmath $\sigma$} \cdot \hat{\bf r} \,
g(r) \end{array} \right)~\chi_m^{[1/2]} \ .
\end{equation}
This yields
\begin{equation}
q(x)~=~{M \over \pi^2}~\int_{k_{min}}^\infty \!\!\!d|{\bf k}|
|{\bf k}|~ {\phi_2(|{\bf k}|) \over \phi_3(0)}
{}~\left[ I_0(|{\bf k}|)^2~+~2 I_0(|{\bf k}|) I_1(|{\bf k}|) \cos\theta_k
{}~+~I_1(|{\bf k}|)^2 \right]
\end{equation}
for the flavor and spin averaged quark distribution function. The term
$\cos\theta_k$ is fixed via Eq. (17), $k_{min}$ is given in Eq.
(18), and with
\begin{mathletters}
\begin{eqnarray}
I_0(|{\bf k}|)~&=&~\sqrt{\pi} \int dr~r^2~f(r)~j_0(|{\bf k}|r) \ , \\
I_1(|{\bf k}|)~&=&~\sqrt{\pi} \int dr~r^2~g(r)~j_1(|{\bf k}|r) \ .
\end{eqnarray}
\end{mathletters}

\figs{11.75}{3.0}{0.3}{-2.1}{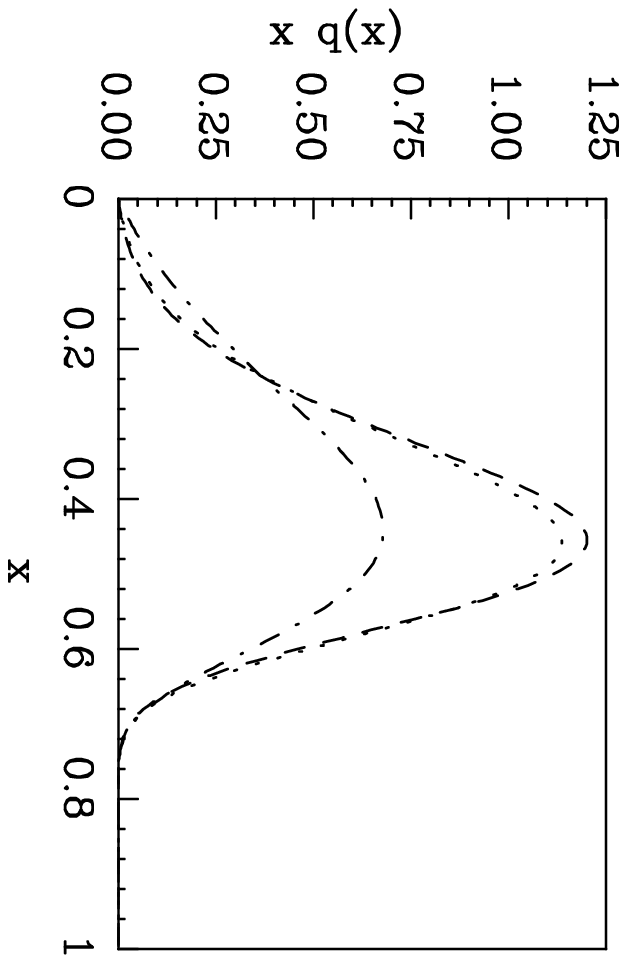}{The spin and flavor averaged
quark distribution in the nucleon is shown as a
function of the struck quark's momentum fraction.
Results are displayed for the CDM
for the parameter sets $f=3$ (dashed line) and $f=\infty$
(dotted line), and for the MIT bag model with a bag radius of $R=1$
fm (dot-dashed line).}

In Fig. 4, we show $x \, q(x)$ for the two parameter
sets and for the MIT bag model
for a bag radius of $R=1$ fm and setting $\epsilon = {M \over 4}$ in
accordance with the bag virial theorem.  Note that due to the
truncation of the sum over all possible intermediate
spectators after only
a single di-quark state, the quark distribution functions
do not satisfy the normalization condition, i.e.,
\begin{equation}
\int_0^1 dx~q(x)~\ne~1 \ .
\end{equation}
The violation is much larger for the MIT bag model ($\approx 17$\%)
than for the chromo-dielectric soliton model ($\approx 2$\%), and it
is usually cured by adding a phenomenological term, $\sim (1-x)^7$,
which is supposed to parameterize the contribution from four-particle
intermediate states \cite{signal88,bate92}.
That term is then
scaled such that the normalization requirement is fulfilled.
As we are only interested in a qualitative discussion,
and as the deviation is anyhow minute for the CDM, we do not adopt
this prescription.  With this, we restrict ourselves to a pure valence
quark picture and neglect all ``sea" contributions.

We see from Fig. 4, that the MIT bag model leads to a wider light-cone
momentum distribution.  This reflects the effect of the sharp cut-off
at the surface of the MIT bag,
which, in turn, yields momentum-components in the quark wave function
that extend to very high momenta and to a broad peak for
$x \, q(x)$.  Note that the location of the peak
is determined by the mass of the spectator state through
$k_{min}$ of Eq. (18), which suggests that $q(x)$ is maximal for
$x \approx 1 - M_S/M$.

To compare our results with experimental data,
the calculated quark distribution functions are evolved from the
relatively low scale, $\mu < 1$ GeV, at which the
bag model is expected to be a reasonable approximation
to non-perturbative QCD, to the experimental momentum scale,
$Q^2 > 5$ GeV$^2$, using the non-singlet Altarelli-Parisi evolution
equations \cite{altarelli77}.  The bag model scale is then
fixed by requiring an optimal fit to a recent parameterization
\cite{cteq} of the experimental iso-singlet valence quark
distribution in the nucleon at $Q^2 = 10$ GeV$^2$. We use
next-to-leading-order QCD evolution in the modified minimal
subtraction scheme ($\overline{MS}$)
with $\Lambda_{QCD} = 213 MeV$ for four flavors \cite{koba94}.
The corresponding results are shown in Fig. 5, and we find
$\mu = 0.33$ GeV for the chromo-dielectric soliton model and
$\mu = 0.38$ GeV for the MIT bag model.

\figs{11.75}{3.0}{0.3}{-2.1}{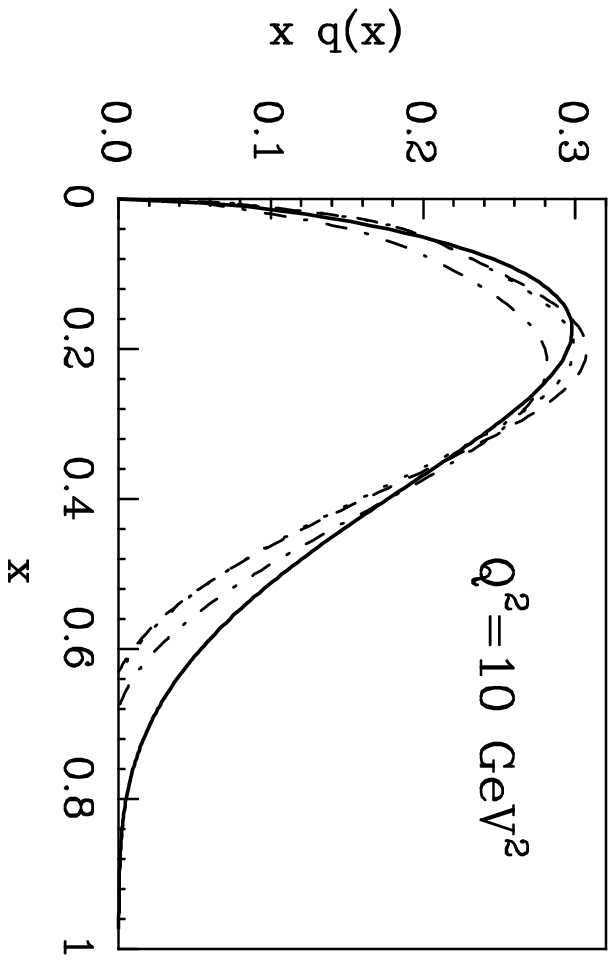}{The spin and flavor averaged
quark distribution
functions of the nucleon are evolved to a momentum scale of $Q^2 = 10$
GeV$^2$.  Results are shown for the chromo-dielectric soliton model
and the MIT bag model, and they are compared with a recent
parameterization of experimental data \protect\cite{cteq}
(solid line).  The remaining
labeling is the same as in Fig. 4.}

Up to $x \approx 0.5$ both models yield a reasonable fit to the
data-based parameterization.  Note that large $x$ at high $Q^2$
correspond to even larger
$x$ before the evolution, i.e., at the scale where the bag model is
valid, and hence to very high quark momenta, for which both
the bag model and the non-relativistic Peierls-Yoccoz
projection become very questionable. The smallness of the bag model
scale is also a concern, due to the question of the applicability of
perturbative QCD.  We comfort ourselves,
however, with the realization that we are ultimately only interested
in a qualitative discussion of a ratio of structure functions, which
should display much weaker $Q^2$ dependence.

We furthermore observe that the peaks of the distributions are
generally higher and/or at larger values of $x$ than the data.  This
deviation, which was also observed in Ref. \cite{bate92}, is more
pronounced for the chromo-dielectric soliton than for the MIT bag
model, and even better fits were obtained for the latter for smaller
bag radii of $R \approx 0.6 \ldots 0.8$ fm \cite{signal88}.
This shows that deep-inelastic scattering data point
towards higher quark momenta and thus smaller bags.
As the parameters of the model,
and hence the size of the bag, were adjusted to fit
standard nucleonic properties, i.e., the mass, the
charge radius and the magnetic moment, latter deviation suggests
that inclusive deep-inelastic scattering does not probe the ``average
nucleon" but is only sensitive to a ``small-size" component in the
nucleon's wave function.  Work in that direction is currently
in progress \cite{koepf95}.

\section{STRUCTURE FUNCTIONS OF SIX-QUARK CLUSTERS}

Previous estimates of the quark light-cone momentum distribution
in six-quark clusters were based on dimensional counting rules
\cite{brodsky74}, valid in the limit $x \to 1$, and on
Regge arguments \cite{kuti71} for low $x$.  Here, we
calculate the quark distribution of six-quark structures
in the framework of a model whose parameters
were adjusted to yield an optimal description of the corresponding
quantity for the nucleon.

The different target mass, $M_{6q}$ instead of $M$, leads to trivial
modifications in the definition of the struck quark's momentum
fraction,
\begin{equation}
x~\equiv~{k'_z \over p'_z}~\approx~{k_0+k_z \over M_{6q}}~
\approx~{M \over M_{6q}}~ x_B \ ,
\end{equation}
and the quark distribution function,
\begin{equation}
q_f^{6q}(x)~=~\sum_S~\delta \left(x-k_+/M_{6q}\right)~| \langle
\, S \, | \, {\hat \psi}_{+,f} (0) \, | \, 6q \, \rangle |^2 \ .
\end{equation}
Employing the same approximations than in the last section, we
truncate this sum after a single spectator state,
which we assume to be a five-quark system that is on its
mass shell with mass $M_S$.

In the following, we restrict ourselves to spherically symmetric
six-quark bags, i.e., to $\alpha = 0$. The six-quark state
can be expressed through terms of the form
$|\sigma^p\pi^{6-p}[f]_O[f']_{TS}\rangle$, with $p \in \{0,2,4,6\}$
and where $[f]_O$ and $[f']_{TS}$ denote Young symmetries.
As we are interested in spin and isospin averaged
quark distributions only, the latter are irrelevant in the
following, and the six-quark cluster can be represented as,
\begin{equation}
| \, 6q \, \rangle~=~a_6 \, | \, \sigma^6 \, \rangle
                  ~+~a_4 \, | \, \sigma^4\pi^2 \, \rangle
                  ~+~a_2 \, | \, \sigma^2\pi^4 \, \rangle
                  ~+~a_0 \, | \, \pi^6 \, \rangle \ ,
\end{equation}
where $|\sigma^p\pi^{6-p}\rangle$ denotes a system with $p$ quarks
in the positive parity state $|\sigma\rangle$, and $6-p$ quarks in
the negative parity state $|\pi\rangle$.  Note that the
single-particle states evolve to $|\sigma\rangle \rightarrow
|s_{1/2}\rangle$ and $|\pi\rangle \rightarrow |p_{3/2}\rangle$
for $\alpha = 0$.  As the terms in
the sum in Eq. (28) add incoherently for the deep-inelastic
process, we can write
\begin{equation}
q_{6q}(x)~=~|a_6|^2 \, q^{6q}_{60}(x)
         ~+~|a_4|^2 \, q^{6q}_{42}(x)
         ~+~|a_2|^2 \, q^{6q}_{24}(x)
         ~+~|a_0|^2 \, q^{6q}_{06}(x)
\end{equation}
for the quark distribution of the six-quark clusters, and
where $q^{6q}_{pq}(x)$ is the distribution function of the
configuration $|\sigma^p\pi^q\rangle$.  The latter obtain
incoherent contributions
from processes where a quark is knocked out of either single-particle
orbital,
\begin{equation}
q^{6q}_{pq}(x)~=~{p \over p+q} \, q^{6q}_{p q \sigma}(x)
              ~+~{q \over p+q} \, q^{6q}_{p q \pi   }(x) \ ,
\end{equation}
where, in generalization of Eq. (23),
\begin{mathletters}
\begin{eqnarray}
q^{6q}_{pq \sigma}(x)~&=&~{M_{6q} \over \pi^2}
\int_{k^{6q}_{min}}^\infty \!\!\!d|{\bf k}| |{\bf k}|~
{\phi_{p-1 \, q}(|{\bf k}|) \over \phi_{pq}(0)}~
{\cal I}^\sigma(|{\bf k}|,\cos\theta^{6q}_k) \ , \\
q^{6q}_{pq \pi   }(x)~&=&~{M_{6q} \over \pi^2}
\int_{k^{6q}_{min}}^\infty \!\!\!d|{\bf k}| |{\bf k}|~
{\phi_{p \, q-1}(|{\bf k}|) \over \phi_{pq}(0)}~
{3\cos^2\theta^{6q}_k + 1 \over 2}~
{\cal I}^\pi   (|{\bf k}|,\cos\theta^{6q}_k)\ ,
\end{eqnarray}
\end{mathletters}
and where
\begin{equation}
{\cal I}^{\sigma , \pi}(|{\bf k}|,\cos\theta^{6q}_k)~=~
I_0^{\sigma , \pi}(|{\bf k}|)^2~+~
2 I_0^{\sigma , \pi}(|{\bf k}|) I_1^{\sigma , \pi}(|{\bf k}|)
\cos\theta^{6q}_k ~+~ I_1^{\sigma , \pi}(|{\bf k}|)^2 \ ,
\end{equation}
with
\begin{mathletters}
\begin{eqnarray}
I_0^{\sigma , \pi}(|{\bf k}|)~&=&~\sqrt{\pi} \int dr~r^2~
f_{\sigma , \pi}(r)~j_{l_{\sigma , \pi}  } (|{\bf k}|r) \ , \\
I_1^{\sigma , \pi}(|{\bf k}|)~&=&~\sqrt{\pi} \int dr~r^2~
g_{\sigma , \pi}(r)~j_{l_{\sigma , \pi}+1} (|{\bf k}|r) \ .
\end{eqnarray}
\end{mathletters}

We employed single quark wave functions of the form
\begin{equation}
\psi_{\sigma , \pi \, \pm {1 \over 2}}({\bf r})~=~
\left(\begin{array}{c}
f_{\sigma , \pi}(r) \\
i \mbox{\boldmath $\sigma$} \cdot \hat{\bf r} \,
g_{\sigma , \pi}(r) \end{array} \right)~
Y_{j_{\sigma , \pi} \, l_{\sigma , \pi}}^{\pm {1 \over 2}}
(\mbox{\boldmath $\Omega$}) \ ,
\end{equation}
where $j_\sigma=1/2$, $j_\pi=3/2$, $l_\sigma=0$ and $l_\pi=1$, and
with the Hill-Wheeler overlap kernels,
\begin{equation}
\phi_{pq}(|{\bf k}|)~=~4\pi \int_0^\infty dR~R^2 j_0(|{\bf k}|R)~
T^p_\sigma(R)~T^q_\pi(R) \ ,
\end{equation}
where
\begin{equation}
T_{\sigma , \pi}(R)~=~\int d^3{\bf r}~
\psi^\dagger_{\sigma , \pi \, \pm {1 \over 2}}({\bf r})~
\psi_{\sigma , \pi \, \pm {1 \over 2}}({\bf r}+{\bf R}) \ .
\end{equation}
The term $\cos\theta^{6q}_k$ is given by
\begin{equation}
\cos\theta^{6q}_k~=~
{\sqrt{M_S^2 + |{\bf k}|^2} - M_{6q}(1-x) \over |{\bf k}|} \ ,
\end{equation}
and the lower limit in the integration over the struck quark's
three-momentum is
\begin{equation}
k^{6q}_{min}~=~\left| {M_{6q}^2(1-x)^2 - M_S^2 \over 2M_{6q}(1-x)} \right| \ .
\end{equation}

In Fig. 6 we display the quark light-cone distributions in the
six-quark clusters which we obtain when including different
parts of the one-gluon-exchange interaction when diagonalizing
the effective Hamiltonian.  In particular, in calculation (I) the OGE
was left out altogether, in calculation (II) only the color-magnetic
hyperfine interaction was considered, and in calculation (III) also
the color-electrostatic force was included.  The distribution
functions are again evolved from the bag model scale, $\mu = 0.33$
GeV, to the experimental scale, $Q^2 = 10$ GeV$^2$.
The quark distributions, $x \, q_{6q}(x)$, are shown as
functions of the Bjorken scaling variable, $x_B$, identifying
the struck quark's momentum fraction through $x={M \over M_{6q}} x_B$,
and they are compared with the corresponding quantity for the nucleon,
$x \, q(x)$, for which $x=x_B$.  For clarity, we restrict ourselves
to one parameter set only.

\figs{11.75}{3.0}{0.3}{-2.2}{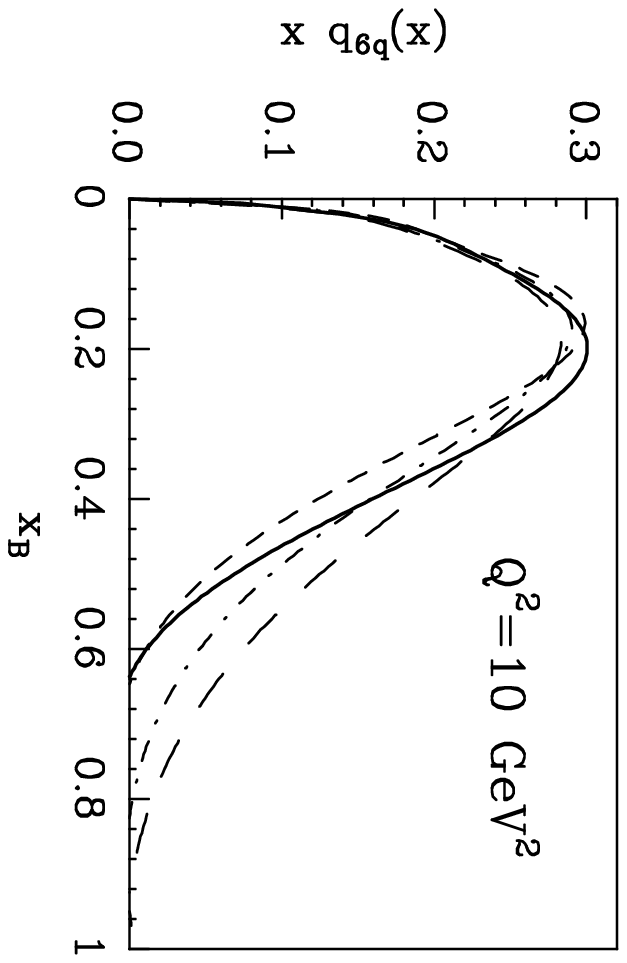}{The spin and flavor averaged
quark distribution in the six-quark clusters, evolved to a
momentum scale of $Q^2 = 10$ GeV$^2$, is shown in comparison with
the nucleon's light-cone distribution (solid line). Depicted
are a calculation where the one-gluon-exchange was left out altogether
(short-dashed line), a calculation where only the color-magnetic
hyperfine interaction was included (dot-dashed line), and
a calculation where also the color-electrostatic force was accounted
for (long-dashed line). Results are shown for (TS)=(10) and
$f=\infty$.}

For the mass of the six-quark state, $M_{6q}$, we write
\begin{equation}
M_{6q}~=~2M~+~V_{ad}(\alpha=0) \ ,
\end{equation}
where $V_{ad}(\alpha=0)$ is the energy difference of a
spherically symmetric six-quark bag and two non-interacting nucleons,
which is $-338$ MeV for calculation (I),
$-69$ MeV for calculation (II) and $+175$ MeV for calculation (III).
The mass of the spectator is set to
\begin{equation}
M_S=M_{6q} -\epsilon \ ,
\end{equation}
where $\epsilon$ is the struck quarks's single particle energy, which
takes the values $272$ ($413$) MeV for the $|\sigma\rangle$
($|\pi\rangle$) state.

We observe that only if the one-gluon-exchange interaction
is neglected, the distribution of the quarks in the
six-quark clusters is shifted to lower $x$, as has been assumed
frequently in the literature
\cite{pirner81,brodsky74,carlson83,lassila88}.
As already pointed out in Sec. III,
the reason for that softening of the quarks' momenta is
the increase of the confining volume.  However, when the
one-gluon-exchange interaction is included, the channel-coupling
it generates
leads to an admixture of higher configurations, and hence to
a hardening of the quarks' average momentum, which competes with the
aforementioned shift to lower $x$.  As can be seen from Fig. 6,
this competition persists in the deep-inelastic domain, and the
conclusions we draw for the non-relativistic quark momentum
distribution apply to the light-cone distributions as
well.  This effect is even more pronounced
if also the color-electrostatic part of the OGE
interaction is considered.

To further elaborate that point, we plot in Fig. 7 the
``structure function ratio" \cite{carlson83}
\begin{equation}
R(x_B)~=~(1-f)~+~f \, {x \, q_{6q}(x) \, \Big|_{x={M \over M_{6q}}x_B}
\over x \, q(x) \, \Big|_{x=x_B}} \ ,
\end{equation}
which has been presented in the literature \cite{carlson83,lassila88}
for a quantitative description of the EMC effect in the
framework of the quark cluster model \cite{pirner81}, together
with experimental deep-inelastic nuclear
cross-section ratios, $F_2^{Fe}(x_B)/F_2^{D}(x_B)$,
of iron over deuterium \cite{benve87,bodek83}.

\figs{11.75}{3.0}{0.3}{-2.2}{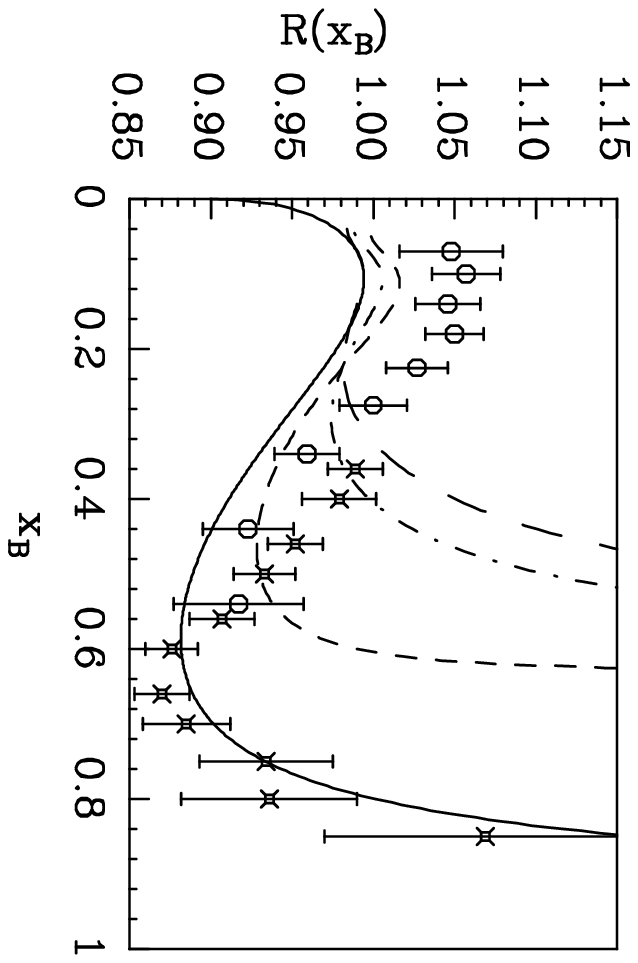}{The structure function ratio
at $Q^2 = 10$ GeV$^2$ is shown in comparison with experimental data
for $F_2^{Fe}(x_B)/F_2^{D}(x_B)$ from Ref. \protect\cite{benve87}
(circles) and Ref. \protect\cite{bodek83} (crosses).  Also shown is
a quark cluster model prediction from Ref. \protect\cite{lassila88}
(solid line).  The remaining labeling is the same as in Fig. 6.}

The quark cluster model is based on the assumption that in finite
nuclei there exists a non-negligible probability, $f$, for the
nucleons to break up into six-quark structures \cite{pirner81}.
If the quark's light-cone distribution in the six-quark clusters
is different from that in a free nucleon, also the
nuclear structure functions, $F_2^{A}(x_B)$, will be different from
their nucleonic counterpart, with
$R(x_B) \approx F_2^{A}(x_B) / F_2^{D}(x_B)$
\cite{carlson83,lassila88}.
As suggested in Refs. \cite{carlson83} and \cite{lassila88}, we use a
probability of $f=0.30$ for the formation of six-quark clusters in the
Fe nucleus.

If we neglect the one-gluon-exchange, the quark
distribution in the six-quark structures is shifted to lower $x$
through the larger confinement volume, and we observe a ``EMC like"
behavior of the ratio $R(x_B)$, as can be seen from the short-dashed
line in Fig. 7 which qualitatively follows the data.
However, if the one-gluon-exchange
is included, the softening of the quark distributions is
cancelled by the admixture of higher and harder
configurations, and the respective ratios $R(x_B)$
are very different from the data.  This is especially the
case if we include the color-electrostatic share of the
one-gluon-exchange, which had proven to be crucial
for a correct description of the short-range
repulsive core of the $N$-$N$ interaction.

\section{CONCLUSIONS}

We have continued our study of the two-nucleon system in terms of
quark degrees of freedom, shifting the focus of our attention from
the adiabatic potential to the quark-substructure.
Employing a relativistic quark bag model, which yields spatial as
well as color confinement, we have constructed six-quark states
which are confined in a deformed bag-like mean field through an
effective non-linear interaction with a self-consistently determined
scalar background field. The corresponding Hamiltonian includes
also quark-quark interactions generated through one-gluon-exchange,
and when evaluating the gluonic propagators, which mediate that
interaction, the inhomogeneity and deformation of the dielectric
medium were taken into account.
Six-quark molecular-type configurations were generated
as a function of deformation, and the effective Hamiltonian was
diagonalized in a coupled channel analysis and while accounting for
different shares of the one-gluon-exchange interaction.

The symmetry structure and the quarks' non-relativistic momentum
distribution in the six-quark state were
discussed, and it was shown, that there
exists an interesting competition between a softening of the quarks'
momenta through the increase in the size of the confining volume,
and a hardening via the admixture of higher symmetry configurations.

Then, the spin and isospin averaged deep-inelastic structure
functions of the nucleon as well
as the six-quark configurations were evaluated following the
formalism presented by the Adelaide group.
In detail, the twist-two
light-front quark distribution functions were calculated from the bag
model wave functions including a Peierls-Yoccoz projection to
cure the broken translational invariance.
The quark distribution functions were
then evolved from the relatively low scale, at which
the bag model is expected to be a reasonable approximation to
non-perturbative QCD, to the experimental momentum scale, using the
non-singlet Altarelli-Parisi evolution equations, and
the bag model scale was adjusted so that the evolved quark
distribution functions agree well with experimental data. The
corresponding light-cone distributions of the six-quark structures
were then predicted, and finally a ``structure function
ratio", proposed previously in the framework of the quark cluster
model, was presented.

It was shown that this ratio displays the typical behavior that
characterizes the EMC effect only if the one-gluon-exchange
interaction is neglected.  However, the underlying shift of the
six-quark distribution functions to lower $x$, which is usually
assumed in investigations in that realm, is cancelled completely by
the admixture of higher and harder configurations  if the
one-gluon-exchange interaction is included.  As the latter was
essential in reproducing the short-range repulsive core of the
adiabatic $N$-$N$ potential, the na\"\i ve picture of a softening of
the quark's distribution functions in the six-quark clusters  due
to the increase of the confining volume, which was put forward
e.g., in the framework of the quark cluster model, is put into
question.

\acknowledgments{We wish to thank J.M. Eisenberg, L.L. Frankfurt,
Fl. Stancu and S. Pepin for many useful discussions.  We also thank
R. Kobayashi, M. Konuma and S. Kumano as well as the CTEQ
Collaboration for providing us with their FORTRAN codes.  This work
was supported in part by the U.S. Department of Energy, and by the
MINERVA Foundation of the Federal Republic of Germany.}


\end{document}